\documentclass[twocolumn]{aa}
\usepackage[colorlinks, citecolor=cyan]{hyperref}
\usepackage{cleveref}
\usepackage{natbib}
\bibpunct{(}{)}{;}{a}{}{,} 
\usepackage{graphicx}
\usepackage{txfonts}
\usepackage{soul}
\usepackage{color, xcolor}
\usepackage{CJKutf8}
\usepackage{orcidlink}

\soulregister\cite7
\begin{document}
 \title{Diverse origins for non-repeating fast radio bursts: Rotational radio transient sources and cosmological compact binary merger remnants}
   \subtitle{}

   \author{Zi-Liang Zhang \begin{CJK*}{UTF8}{gbsn} (张子良) \end{CJK*} \inst{1,2} \orcidlink{0000-0002-9110-4336},
            Yun-Wei Yu \begin{CJK*}{UTF8}{gbsn} (俞云伟) \end{CJK*} \inst{1,2} \orcidlink{0000-0002-1067-1911},
            Xiao-Feng Cao \begin{CJK*}{UTF8}{gbsn} (操小凤) \end{CJK*} \inst{3}
                    }

   \institute{Institute of Astrophysics,
   			  Central China Normal University, Wuhan 430079, China\\
              \email{yuyw@ccnu.edu.cn}
              \and
              Key Laboratory of Quark and Lepton Physics (Ministry of Education), Central China Normal University, Wuhan 430079,
China
\and
School of Physics and Electronic Information, Hubei University of Education, Wuhan 430205, China\\
 \email{caoxf@mails.ccnu.edu.cn}
         \\
             }

   \date{}

\abstract{A large number of fast radio bursts (FRBs) detected with the CHIME telescope have enabled  investigations of their energy distributions in different redshift intervals, incorporating\ the consideration of the selection effects of CHIME. 
As a result, we obtained a non-evolving energy function (EF) for the high-energy FRBs (HEFRBs) of energies $E\gtrsim2\times0^{38}$ erg, which takes the form of a power law with a low-energy exponential cutoff. 
On the contrary, the energy distribution of the low-energy FRBs (LEFRBs) obviously cannot be described by the same EF. 
Including the lowest dispersion measure (DM) samples, the LEFRBs are concentrated towards the Galactic plane and their latitude distribution is similar to that of Galactic rotational radio transients (RRATs). 
These indications hint that LEFRBs might compose a special type of RRATs, with relatively higher DMs and energies (i.e., $\sim10^{28-31}$ erg for a reference distance of $\sim10$ kpc if they belong to the Milky Way). 
Finally, we revisit the redshift-dependent event rate of HEFRBs and confirm that they could be produced by the remnants of cosmological compact binary mergers. }

\keywords{fast radio bursts}

\maketitle
\section{Introduction} 
\label{sec:intro}
Fast radio bursts (FRBs) are short and intense radio transients of unusual dispersion measures (DMs), which are a general indication of cosmological distance and large energy releases (e.g., \citealt{Lorimer2007,Keane2012,Thornton2013}
; see \citealt{Zhang2020Nature} and \citealt{Petroff2022} for recent reviews). 
The energy requirement and millisecond duration of FRBs hint at the likelihood that they may result from violent activities of compact objects, in particular,  highly magnetized neutron stars \citep[NSs;][]{Popov2010,Kulkarni2014,Lyubarsky2014,Geng2015,Dai2016,Wang2016,Gu2016,Katz2016a,Connor2016,Cordes2016,Lyutikov2017,Zhang2017}. 
Strong support for the NS activity model had  been further provided by the discovery of some repeating FRBs  \citep[{ e.g.,}][]{Spitler2016Nature,Tendulkar2017,Chatterjee2017Natur,CHIME2019a,CHIME2019b}.
The persistent radio counterpart of repeating  FRB 20121102A  even enabled us to constrain the environment and age of the NS \citep{Cao2017b,Kashiyama2017,Metzger2017,Dai2017,Michilli2018Nature}. 
Recently, the observation of FRB 20200428 and FRB 20221014 from the Galactic magnetar SGR 1935+2154, which were temporally associated with an X-ray burst \citep{Bochenek2020,CHIME2020,HXMT2021,2022ATel15681....1D,2022ATel15682....1W}, offered smoking-gun evidence of the NS origin of some FRBs.
Nevertheless, it is unclear whether these Galactic FRBs can represent their cosmological cousins, because their energy releases are  significantly lower than the cosmological ones.
    
No matter whether FRBs are cosmological or Galactic, it is always important to ask where the FRB NSs are located and what they originate from. 
Magnetar SGR 1935+2154 is associated with the supernova remnant G57.2+0.8 \citep{Gaensler2014,Kothes2018,Zhong2020,Zhou2020}, which suggests it was born from a core-collapse supernova event.
However, the repeating FRB 20200120E had been localized in an old globular cluster (GC) in the nearby galaxy M81 \citep{Kirsten2022Nature}, indicating an association with old stellar population. 
Because of the high stellar densities, various high-energy sources are ubiquitous in GCs, including X-ray binaries, millisecond pulsars (MSPs), cataclysmic variables, and compact binary mergers. 
Therefore,  the NS producing FRB 20200120E could be an accreting NS, a MSP, a young NS formed from accretion-induced collapse of white dwarf (WD) or a merger product of compact binaries \citep{Kremer2021,Lu2022}. 
Although the observations of SGR 1935+2154 and FRB 20200120E are robust, we still cannot use these solitary observations to judge the origin of most other FRBs, which might be produced though other channels.
The diversity in terms of the origins  of  FRBs has also been supported by the identification of the host galaxies of 19 FRBs, which  points toward a wide range of galaxy types. 
For repeaters, they are sometimes close to star-forming regions \citep{Bassa2017,Marcote2020,Fong2021,Ravi2021,Piro2021,Nimmo2021} and sometimes largely deviate from star-forming regions \citep{Tendulkar2021}.
On the other hand, non-repeaters are usually found from low star-forming galaxies and even outskirts of host galaxies \citep{Heintz2020,Mannings2021,Bhandari2022}.

In addition to direct observational inferences, statistical studies of FRBs can provide complementary constraints on FRB origins \citep{Yu2014,Bera2016,Caleb2016,Katz2016,Li2017,Lu2016,Oppermann2016,Vedantham2016,Fialkov2017,Lawrence2017,Cao2018b,Macquart2018,Zhang2021,James2022}.
For non-repeating FRBs, \cite{Yu2014} and \cite{Cao2017a,Cao2018a} first invoked redshift-dependent rates in the statistics and investigated the possible relationship of these rates with the cosmic star formation rates (CSFRs). 
It was found that an extra redshift evolution should be involved, which may be indicative of a time delay between the FRB production and the star formation. 
A possible explanation of this result is that the NSs producing FRBs are formed from compact binary mergers\footnote{Please note that what are produced by the merger events are the FRB NSs rather than the FRB emission, as firstly suggested by \cite{Cao2018a}.} and thus the time delay is determined by binary evolution and the gravitational wave decay of orbits.
This explanation is also consistent with the offsets of FRBs from the center of their host galaxies. Thanks to a concerted global effort in recent years, especially with respect to the Canadian Hydrogen Intensity Mapping Experiment Fast Radio Bursts project (CHIME/FRB), the observed number of FRBs has been rising swiftly and enabling more detailed statistical works \citep{Rafiei-Ravandi2021,Chawla2022,Pleunis2021,Josephy2021,Zhang2022,Hashimoto2022,Bhattacharyya2022}. 

Current statistical studies for non-repeating FRBs are always based on the assumption of a single origin for all of them.
However, this assumption may be not the reality. 
Therefore, in the next section, we revisit the energy distribution of the CHIME FRBs by separating them into several different redshift ranges and taking into account the selection effects of CHIME.
A difference between the low- and high-energy samples can be shown. 
Then, in view of the diversity of FRB origins, we analyze the possible different origins of these two sub-samples in Sections 3 and 4, respectively.
A summary and conclusions are given in Section 5.

\begin{figure}[htbp]
\centering
\includegraphics[width=0.48\textwidth,trim=10 50 20 40,clip]{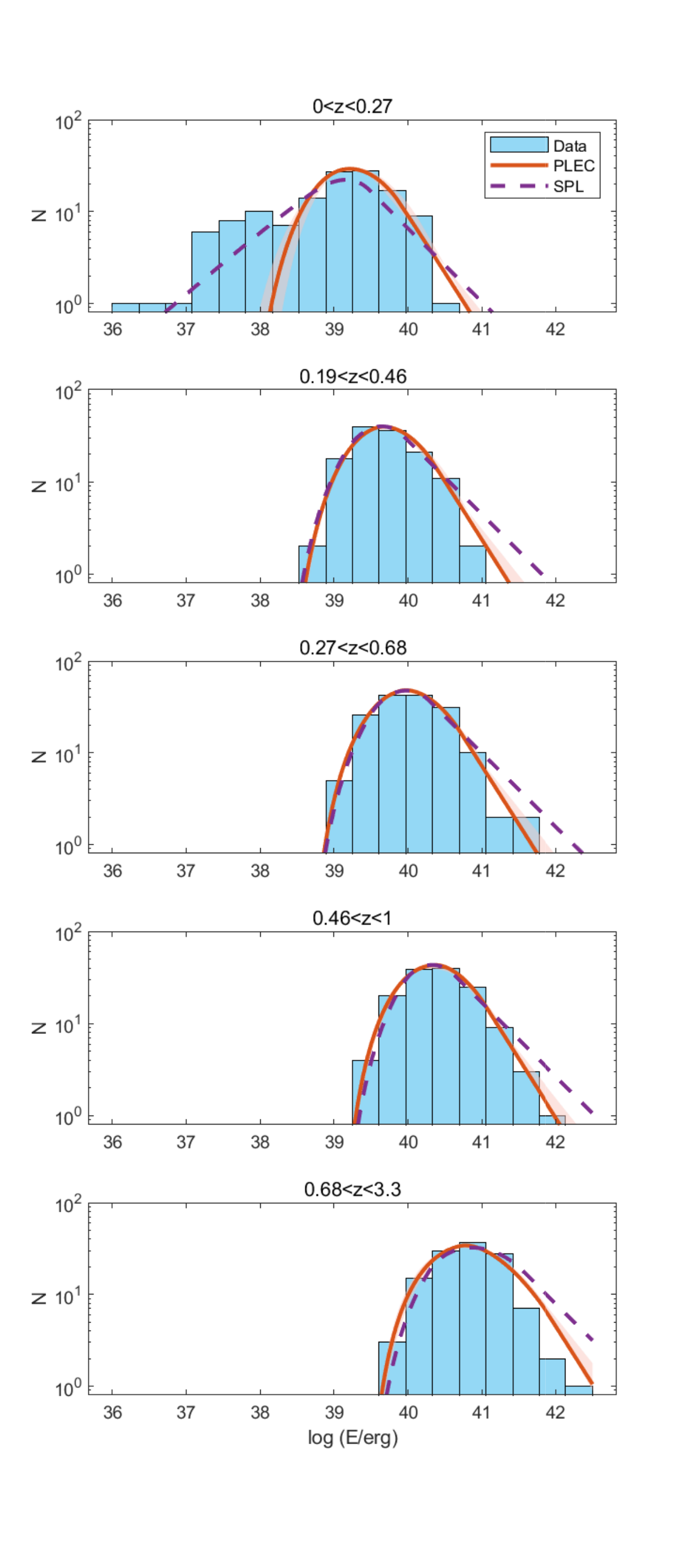}
\caption{
Energy distributions of FRBs in different redshift intervals as labeled. 
The red solid lines and purple dashed lines represent the best fits to the energy distributions for $E>2\times10^{38}$ erg with Equation \eqref{energy fitting} 
The shaded area around the lines represents the 95\% confidence range of the fits. The corresponding parameter values are listed in Table \ref{table alpha}.
}
\label{fig energy function}
\end{figure}

\section{Energy distribution of non-repeating FRBs}
\subsection{CHIME/FRB data} \label{The observational data}
The data from the first CHIME/FRB catalog\footnote{\url{https://www.chime-frb.ca/catalog}} contain 536 events \citep{CHIME2021}. 
Among them, 62 events belonging to 18 repeating FRBs  are not considered in this paper.
We further abandoned the events of  ``Fluence = 0" and the ones noted by ``excluded\_flag" field, which were detected during the software upgrade. 
As a result, 434 events remained in our sample.

By assuming a cosmological origin of all selected FRBs, we can decompose their DMs as follows:
\begin{equation} \label{total DM}
\rm DM  = DM_{MW} + DM_{halo} + DM_{IGM} + \frac{DM_{host}}{1+\emph{z}}+{DM_{\rm sr}\over 1+\emph{z}},
\end{equation}
where $\rm DM_{MW}$ and $\rm DM_{halo}$ represent the contribution from the Milky Way and its halo, $\rm DM_{IGM}$ is related to the column density of intergalactic medium (IGM) on the line of sight, from which we can estimate the redshifts of the FRBs. Next, $\rm DM_{host}$ and $\rm DM_{sr}$ are determined by the host galaxies and the source objects of FRBs, respectively. 
From the CHIME/FRB catalog, we can directly find the value of $\rm (DM - DM_{MW})$ for the FRBs, where the value of $\rm DM_{MW}$ is given with the YMW16 electron density model \citep{YMW16}.
Following previous works \citep{Zhang2021,Zhang2022,Hashimoto2022,Qiang2022}, we further adopt $\rm DM_{halo}= 30\ pc \ cm^{-3}$ \citep{Dolag2015,Prochaska2019} and $\rm DM_{host}= 107\ pc \ cm^{-3}$ \citep{Li2020}. 
Finally, ignoring the DM contribution of the source objects, we ascribe the remaining DM all to the contribution from the IGM, which is expressed as \citep{Deng2014,Macquart2020}:
\begin{equation}
\mathrm{DM}_{\mathrm{IGM}}(z)= 872 \int_{0}^{z} \frac{1+z}{\sqrt{\Omega_{\rm M}(1+z)^3+\Omega_{\Lambda}}} \mathrm{d}z \ \rm   pc \ cm^{-3},\label{DMigm}
\end{equation}
where the cosmological parameters are taken as $\Omega_{\rm M}=0.315$ and $\Omega_{\Lambda}=0.685$ \citep{Planck2018}. 
It should be noticed that for 20 FRBs of relatively low DM, the above calculations would lead to negative values for their $\mathrm{DM}_{\mathrm{IGM}}$, which is clearly incorrect. 
The reason of these results could be due to an overestimation of the DM contribution of their host galaxies or even because these FRBs are not, in fact, cosmological. 
Thus, in order to avoid this uncertainty, we excluded these 20 low-DM FRBs from our statistics in this section. 

From Eq. (\ref{DMigm}), we can derive the redshifts of the selected FRBs and then calculate the isotropically-equivalent energies by \citep{Zhang2018}:
\begin{equation} 
    E = \frac{1 }{1+z}4\pi d^2_{\rm L} F_{\nu}\nu_{\rm c},  \label{Eiso}
\end{equation}
where the luminosity distance is given by: 
\begin{equation}
d_{\rm L}(z)={ c(1+z)\over H_{0}}\int_{0}^{z} {1\over\sqrt{\Omega_{\rm M}(1+z)^3+\Omega_{\Lambda}}} \mathrm{d}z,
\end{equation}
with $H_0 = 67.4 \ \rm km \ s^{-1} \ Mpc^{-1}$, $F_{\nu}$ is the specific fluence of the FRBs and $\nu_{\rm c} = 600 \rm \ MHz$ is the central frequency of the CHIME observational band. 
A more exact estimation of the FRB energy requires a good understanding of their emission spectrum \citep{Houben2019,Beniamini2020}, which can help to correct the above energy into a wider frequency range. 
However, since the spectrum is actually unclear, here we can only use the approximation given by Eq. (\ref{Eiso}), which is generally considered to be acceptable \citep{Zhang2018}.

By using the obtained FRB data of $(z,E)$, we can constrain the energy function (EF) $\Phi(E)$ and redshift-dependent event rates $\dot{R}(z)$ of these FRBs, although these two factors are highly coupled with each other, as shown in previous works \citep[e.g.,][]{Cao2017a,Cao2018a,Luo2020,Zhang2021,James2022,Qiang2022,Zhang2022}. 
At present, fortunately, the sufficiently high number of CHIME FRBs enables us to separate them into several  narrow redshift ranges. 
Specifically, five redshift intervals are taken as $0<z<0.27$, $0.19<z<0.46$, $0.27<z<0.68$, $0.46<z<1.00$, and $0.68<z<3.30$, which correspond to FRB numbers 100, 130, 160, 141, and 123, respectively. 
Here the adjacent intervals are taken to overlap with each other, in order to get sufficiently large FRB sub-samples for each redshift interval.  
The energy distributions of the sub-samples for different redshift intervals are displayed by histograms in Figure \ref{fig energy function}. 
Because all of the redshift ranges are relatively narrow, a constant FRB rate can be taken approximately when we model these energy distributions. 
As a result, the potential redshift evolution of the FRB rate and even the EF can be obtained directly by comparing the constrained parameter values of different redshift intervals. 
In other words, the degeneracy between $\Phi(E)$ and $\dot{R}(z)$ can be solved naturally using this method.

\subsection{Modeling the energy distributions}
As a direct impression of the energy distributions shown in Figure \ref{fig energy function}, the high-energy part of them all exhibit a simple power-law profile. 
Since this energy range is inclined to be unaffected by the threshold effect of telescope, the observational power-law distribution probably indicates an intrinsic power-law EF of the FRBs, which was indeed usually found in previous works. 
Since the EF must not diverge at low energies, we assume that the power law would be cut off by an exponential function at a characteristic energy of $E_{\rm c}$ as: 
\begin{equation} 
\Phi(E) \propto \left({E\over E_{\rm c}}\right)^{-\alpha}\label{EnFun}\exp\left(-{E_{\rm c}\over E}\right ).
\end{equation}
Then, by introducing a detection efficiency $\vartheta$, the observational energy distributions can be given as:
\begin{eqnarray} \label{energy fitting}
\frac{\mathrm{d}N}{\mathrm{d}E}&\propto&
\Phi(E)\int_{z_{\rm min}}^{z_{\rm max}}\vartheta[F_{\nu}(E,z'),{\rm DM}(z')]
 \frac{\mathrm{d} V}{1+z'},\label{Enerdis}
\end{eqnarray}
where $(z_{\rm min},z_{\rm max})$ correspond to the range of the redshift intervals, $\mathrm{d}V(z)=4\pi d_c(z)^2 cH(z)^{-1}\mathrm{d}z$ is the comoving volume element, $d_c(z)=c\int_0^{z}H(z')^{-1}\mathrm{d}z'$ is the comoving distance, and the factor of $(1+z')$ represents the cosmological dilation for the observer's time. 
The detection efficiency is in principle dependent on both the fluence and DM of FRBs and thus we can express it as follows:
\begin{eqnarray}
\vartheta(F_{\nu},{\rm DM})=\zeta(F_{\nu})\eta(\rm DM). \label{selection}
\end{eqnarray}
On the one hand, \cite{CHIME2021} and \cite{Hashimoto2022} had given
 \begin{eqnarray} \label{DM selction}
    \eta(\textrm{DM})& =& 0.8959[-0.7707(\textrm{log}\  \textrm{DM})^2+ 4.0561(\textrm{log}\  \textrm{DM})\nonumber\\
    && - 5.6291],        
\end{eqnarray}
where the coefficient $0.8959$ is given for the normalization of DM detection efficiency.
On the other hand, as suggested by \cite{Zhang2022}, the threshold of CHIME is very likely to be a ``gray zone," rather than a fix value as assumed previously for simplicity. 
This means, below a threshold fluence $F^{\max}_{\nu, \rm th}$, the detection efficiency should gradually decrease with the decreasing fluence. The detection efficiency can be empirically expressed as:
\begin{equation} 
        \zeta_(F_{\nu}) =
        \begin{cases}
             0,  &  \  F_{\nu} \le \  F_{\nu, \rm th}^{\rm min} \\
           \left[\frac{\log (F_{\nu}/F_{\nu, \rm th}^{\rm min}) }{ \log (F^{\rm max}_{\nu, \rm th} /F_{\nu, \rm th}^{\rm min}) }\right]^3, 
           &\  F^{\rm min}_{\nu, \rm th}<F_{\nu} <  F^{\rm max}_{\nu, \rm th}  \\
            1, &  \  F_{\nu} \ge  \  F^{\rm max}_{\nu, \rm th}
        \end{cases}.\label{TelTheresh}
    \end{equation}
According to the observational data, we can take directly $ F^{\rm min}_{\nu, \rm th}=0.3\rm \ Jy~ms$, while $F^{\max}_{\nu, \rm th}$ is taken as a free parameter.

\begin{table}[htbp]
\centering\caption{Parameters for the EF}
\begin{tabular}{c|ccc}
\hline
\hline
Model&$ F^{\rm max}_{\nu, \rm th} /\rm (Jy \ ms)$ & $\alpha$&$ \textrm{log}\  E_{\rm c} /\rm (erg)$\\
\hline
SPL     & $2.63^{+0.93}_{-0.67}$ & $1.80_{-0.12}^{+0.12}$&$-$\\
\hline
PLEC   &$6.73^{+6.61}_{-3.21}$ & $2.27_{-0.26}^{+0.17}$ & $38.71^{+0.23}_{-0.27}$\\
\hline
\end{tabular}
\label{table alpha}
\end{table}

The fitting to the observational energy distributions of FRBs are presented in Figure \ref{fig energy function} by lines and the corresponding parameters are listed in Table \ref{table alpha}. 
First of all, a single power law (SPL) EF without a cutoff was tried in the fitting, in view of the fact that the low-energy cutoff could be much lower than the minimum observational energy.
However, in this case, it is nearly impossible to reconcile the fittings of the low- and high-energy parts, even if we abandon the samples of lowest energies of $E\lesssim 2\times10^{38} \ \rm erg$, where an obvious number excess appears. 
This difficulty in the SPL model could be reduced only if we consider that the EF including its index can evolve with redshift.
By contrary, the invoking of the low-energy cutoff can easily lead to an united and good fitting of the distributions in all redshift ranges, but except for the low-energy excess.
On the one hand, the power law with an low-energy exponential cutoff (PLEC) can provide an effective description for the EF of high-energy FRBs (HEFRBs) and no extra redshift evolving is needed. 
On the other hand, 
this EF inferred from the HEFRBs can not be extended to the low energy range and the low-energy excess is probably true. 
This may indicate, if all of the non-repeating FRBs still own a common origin, then (i) a second EF component is needed for low-energy FRBs (LEFRBs) and (ii) the suppression of the detection efficiency in the low-energy range is overestimated. 
Nevertheless, even taking these model remedies into account, it is still difficult to account for the abrupt disappearing of the LEFRBs at relatively high redshifts ($z\gtrsim 0.27$). 
Then, a possible explanation is that the LEFRBs might have an origin completely different from the high-energy ones and LEFRBs can only be observed at near distances.

It should be emphasized that the separation line between HEFRBs and LEFRBs, which is set around $2\times10^{38}$ erg here, is actually not strict and somewhat dependent on the assumptions of the values of $\rm DM_{halo}$, $\rm DM_{host}$, and $\rm DM_{\rm sr}$.
Nevertheless, the possible shift of this separation line would not eliminate the particularity of LEFRBs. 

\section{Considering a possible Galactic origin for LEFRBs }       
\begin{figure*}[htb]
\centering
\includegraphics[width=0.7\textwidth,trim=30 75 0 70,clip]{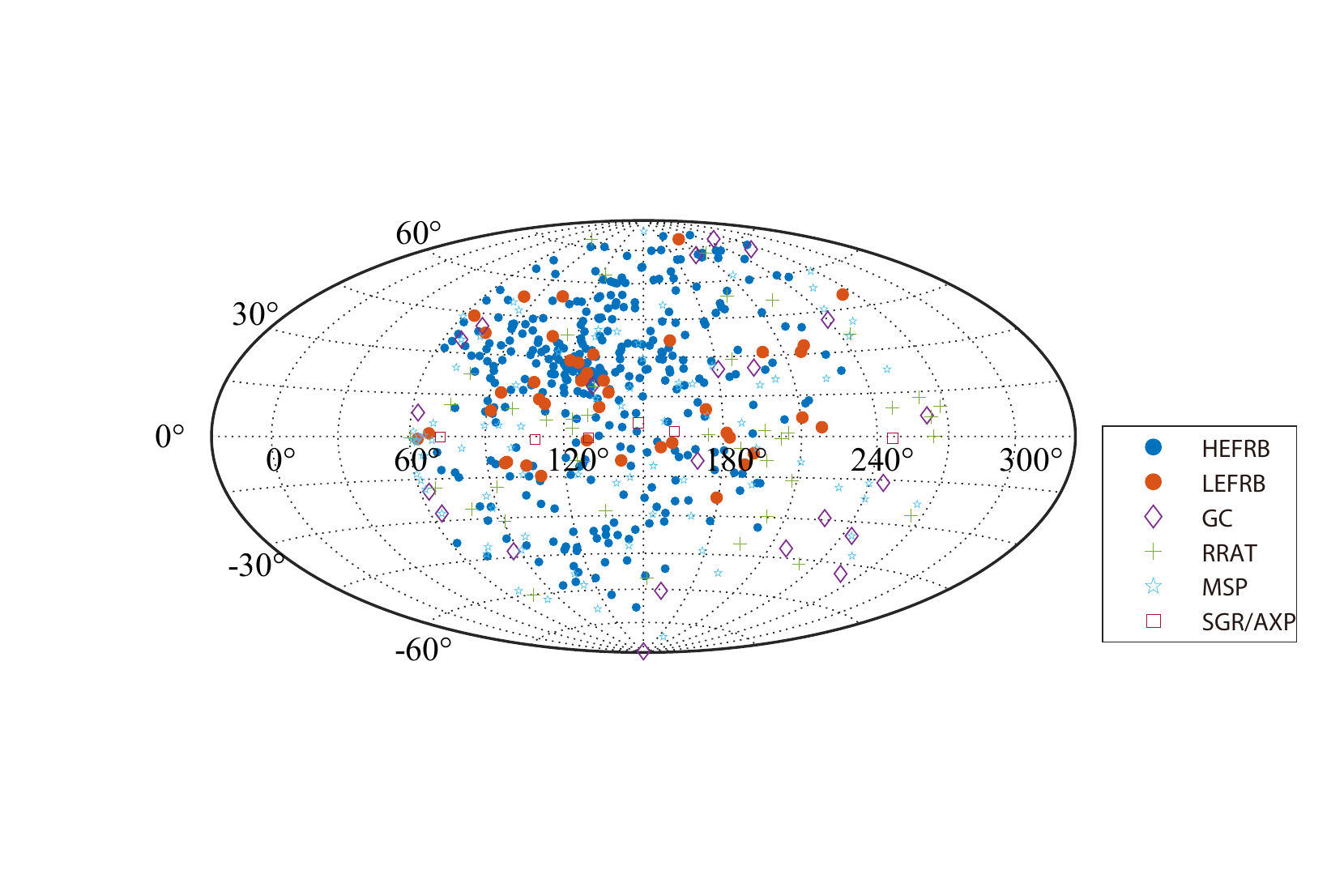}
\caption{Spatial distributions of HEFRBs (blue solid circle) and LEFRBs (red solid circle). 
The Galactic sources including SGRs/AXPs, GCs, MSPs, and RRATs are also represented for comparison, which are taken from \citet{Olausen2014}, \citet{Harris1996}, and \citet{Manchester2005}, respectively. The longitude range is limited within $(60^\circ,270^\circ)$, since outside this range the sensitivity of CHIME is ambiguous.
}\label{sky_distribution}
\end{figure*}

\begin{figure*}[htb]
\centering
\includegraphics[width=0.45\textwidth]{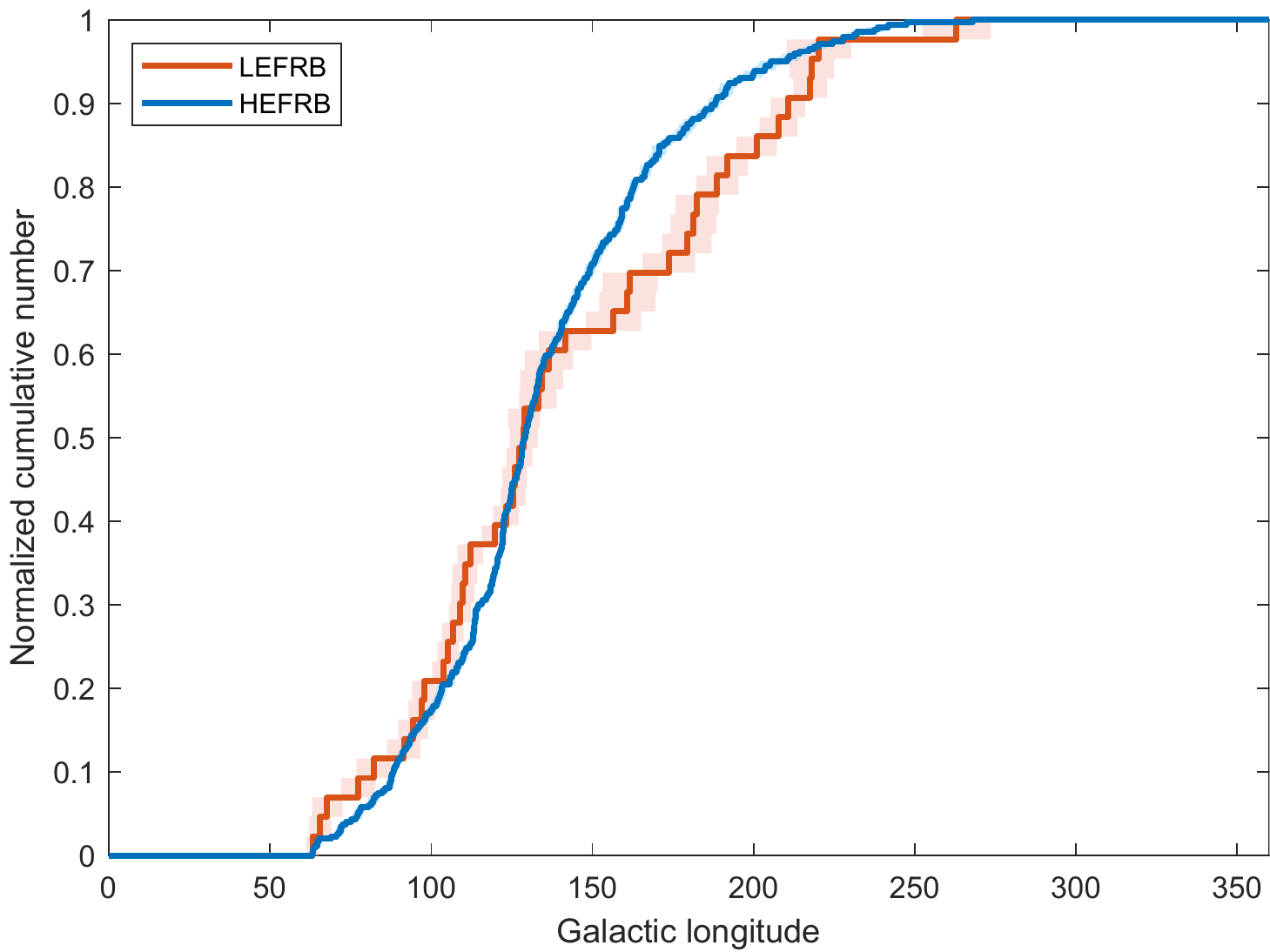}\includegraphics[width=0.45\textwidth]{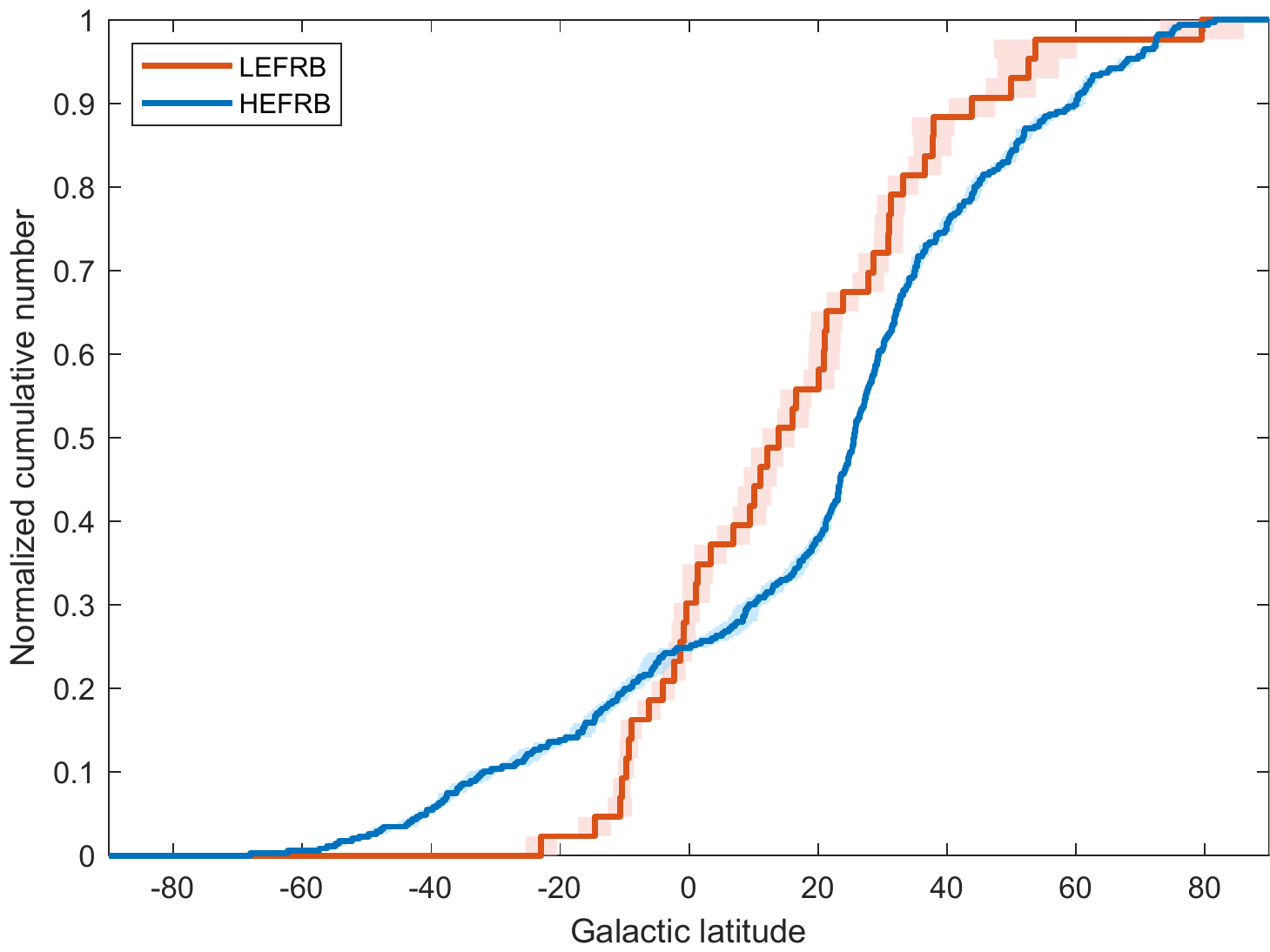}
\caption{Normalized cumulative number of HEFRBs and LEFRBs on the Galactic longitude (left) and latitude (right).
Shaded area around the lines represent the uncertainty interval of sample size effect quantified by bootstrap.
p value of ks test is 0.272 for longitude and 0.014 for latitude.
}
        \label{SDhistogam}
\end{figure*}

In order to clarify the relationship between LEFRBs and HEFRBs, 
we plot their spatial distributions in Figure \ref{sky_distribution}.
Here, besides the 31 LEFRBs defined by their energies $E\lesssim 2\times10^{38}$ erg, we also classify the 20 low-DM FRBs, which are excluded in the above statistics, into our LEFRB class, in view of their potential small distances and low energies. 
As shown in Figure \ref{SDhistogam}, the longitude distributions of  LEFRBs and HEFRBs are generally identical for a $p-$value of $0.272$ returned by the Kolmogorov-Smirnov (KS) test.
However, the latitude distributions with $p_{\rm KS}=0.014$ might not come from the same distribution.

\begin{figure*}[hbt]
\centering
\includegraphics[width=0.8\textwidth]{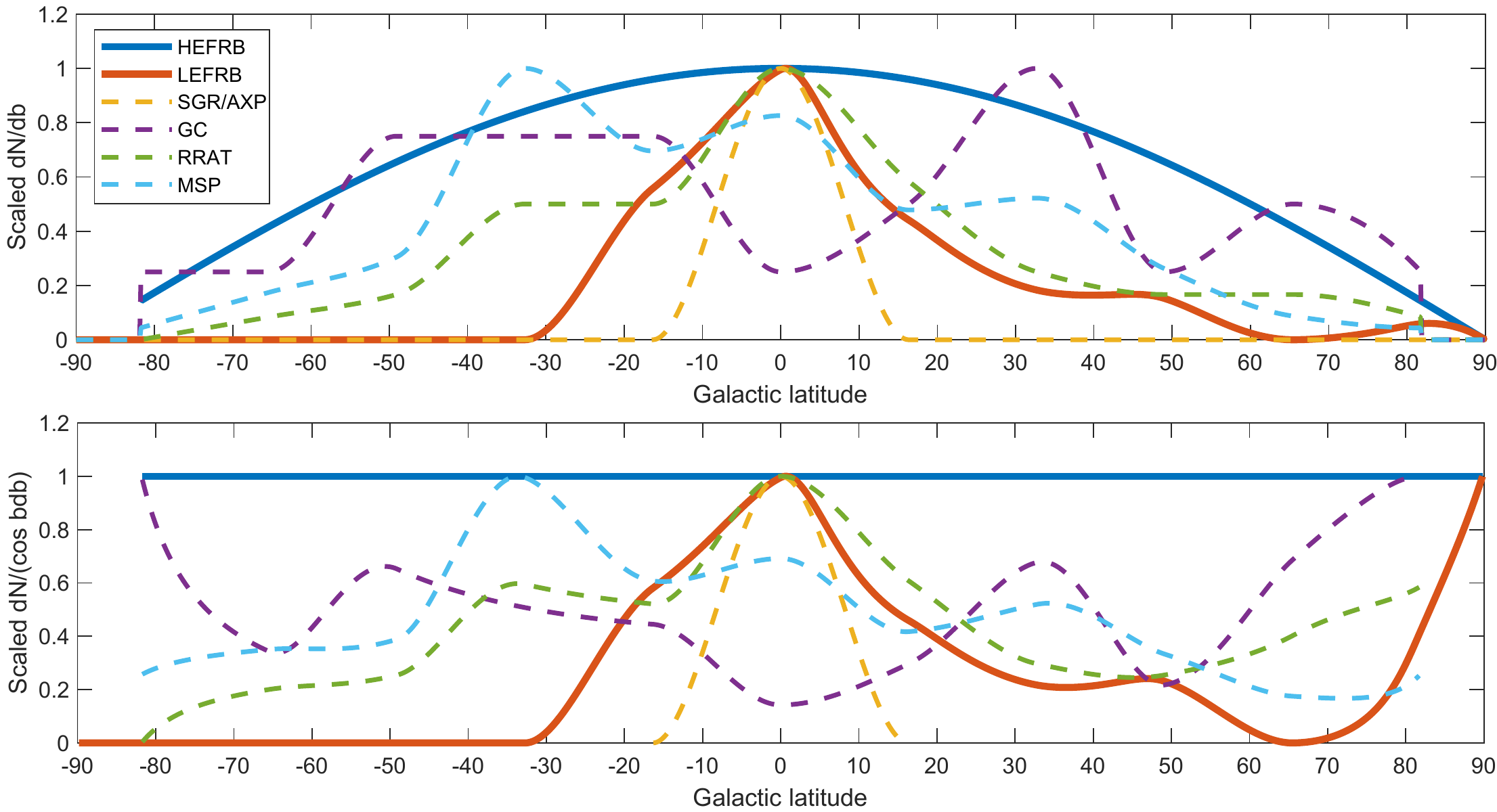}
\caption{Galactic latitude distributions of HEFRB and LEFRB, where the direction-dependence of the detection efficiency of CHIME has been corrected by using Eqs. (\ref{flb}-\ref{LEFRBLatitude}).
The top and bottom panels show the distributions of ${\rm d}N/{\rm d}b$ and ${\rm d}N/(\cos b{\rm d}b)$, respectively. The latitude distributions of Galactic SGRs/AXPs, GCs, MSPs and RRATs are shown for comparison, which are obtained by interpolating the observational histograms. All lines have been scaled to have a peak value of unity. }
\label{Dislatitude}
\end{figure*}

It has been well established that HEFRBs have a cosmological origin, which has been confirmed by the localization of their host galaxies. 
Thus, the intrinsic spatial distribution of HEFRBs is considered to be uniform and the observed spatial distribution of them is a result of the direction-dependence of the detection efficiency of CHIME. 
Following this consideration, we can define:
\begin{equation} \label{flb}
f(l,b)=\frac{{\rm d}N_{\rm HEFRB}^{\rm obs}}{\rho_{\rm HEFRB} \cos{b}{\rm d}b{\rm d}l } ,
\end{equation}
where $\rho_{\rm HEFRB}$ is a constant, $l$ and $b$ are the Galactic longitude and latitude. According to the above expression, we can derive the intrinsic density of the LEFRBs on the celestial sphere by: 
\begin{equation}\label{rhoLEFRB}
\rho_{\rm LEFRB}(l,b)=\left.\frac{{\rm d}N_{\rm LEFRB}^{\rm obs}}{\cos{b}{\rm d}b{\rm d}l}\right/f(l,b).
\end{equation}
In view of the potential axial-symmetry of the spatial distribution, the number distribution of LEFRBs on the Galactic latitude is more concerned, which can be given by:  
\begin{eqnarray}
\frac{ {\rm d} N_{\rm LEFRB} }{{\rm d}b} &\propto& \cos b\cdot\rho_{\rm LEFRB}(l,b),\nonumber\\
&\propto& \cos b {{\rm d} N_{\rm LEFRB}^{\rm obs}/ {\rm d}b\over {\rm d} N_{\rm HEFRB}^{\rm obs}/ {\rm d}b},
\label{LEFRBLatitude}
\end{eqnarray}
while the situation of HEFRBs can be simply written as:
\begin{eqnarray}
\frac{ {\rm d} N_{\rm HEFRB} }{{\rm d}b}&\propto& \cos b.
\end{eqnarray}
Using these expressions, we can plot corrected latitude distributions of LEFRBs and HEFRBs in Figure \ref{Dislatitude} by the solid lines.

The corrected latitude distribution of LEFRBs significantly deviates from the uniform situation and, instead, approaches to be concentrated within the Galactic plane of a variance of $\sigma_{\rm LEFRB}=22^\circ$.
This concentration hints that LEFRBs could originate from the Milky Way. 
Then, for a comparison, we also plot the latitude distributions of some typical Galactic sources in Figure \ref{Dislatitude}, including magnetars (consisting of soft gamma-ray repeaters (SGRs) and anomalous X-ray pulsars (AXPs)), GCs, MSPs, and rotational radio transients (RRATs), which might have potential connections with Galactic FRBs. 
By comparison, although the distributions undulate, which is obviously due to the very limited sample numbers, we can find that the latitude distributions of GCs and MSPs are more diffuse than that of LEFRBs, while the distribution of SGRs/AXPs is too narrow. 
The closest distribution is provided by RRATs, which are also mysterious objects. 

\begin{figure}[htbp]
\centering
\includegraphics[width=0.45\textwidth]{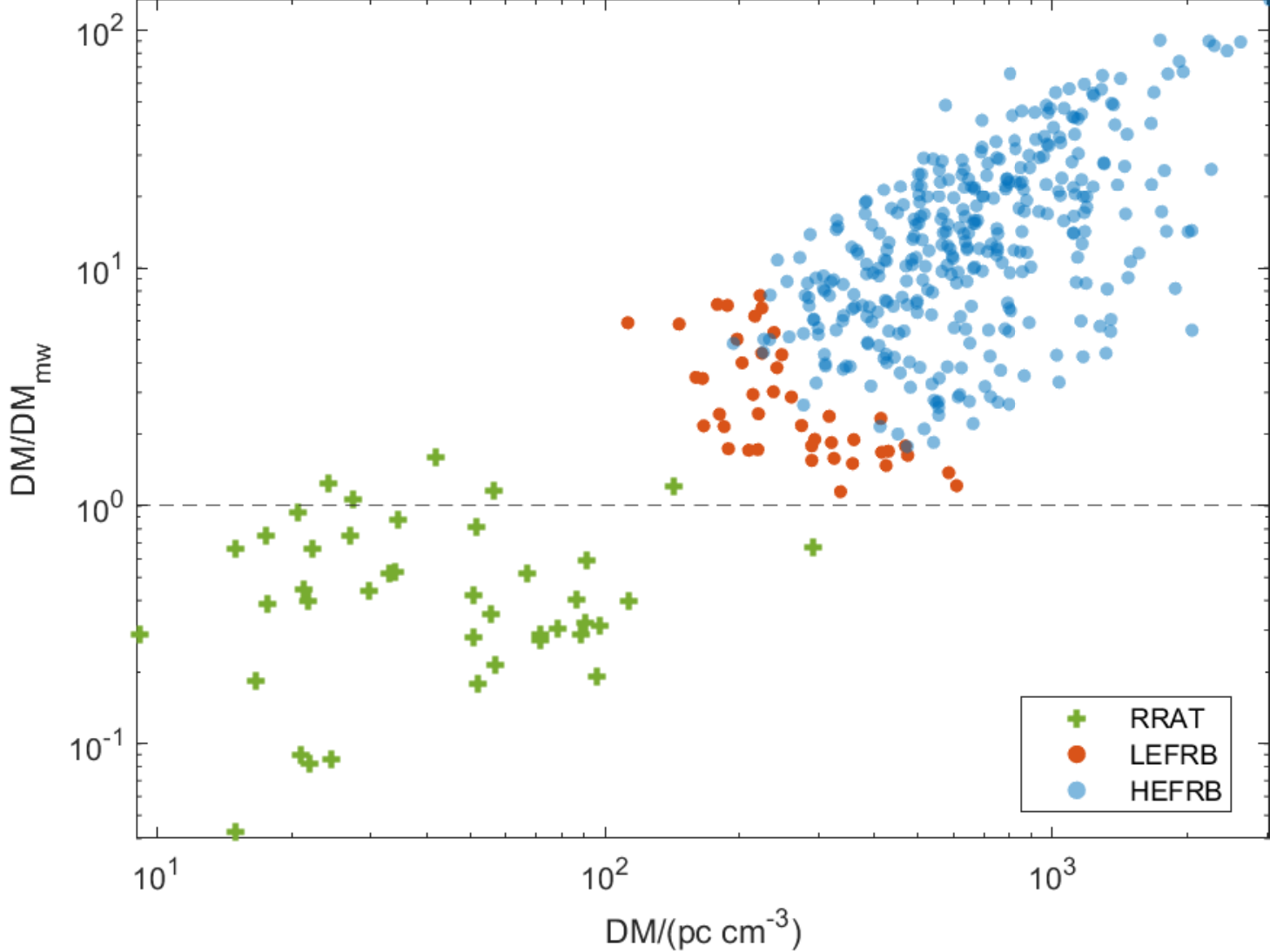}
\caption{Observed $\rm DM$ vs. the ratio of $\rm DM/DM_{mw} $ of LEFRBs and RRATs.}\label{FRB&RRATcomp}
\end{figure}

RRATs are a group of sporadically pulsing sources, which can be classified as a special type of pulsars as they can emit detectable pulses repeatedly \citep{McLaughlin2006,Keane2011,Keane2016c}. 
Nevertheless, there are still a dozen RRATs that have never shown a second pulse, which makes them indistinguishable from FRB pulses in appearance except for their different DMs \citep{Keane2016c}. 
In other word, a ``grey area" exists in the classification of these two types of phenomena \citep{Keane2016c,Rane2017}. 
In principle, it is reasonable to suspect that some single pulses classified as RRATs are actually cosmological FRBs. 
However, this possibility is disfavored by the unsuccessful determination of their host galaxies \citep{Rane2017}. 
Alternatively, it is also of probability that some single pulse events now labelled as FRBs actually belong to the Milky Way and the LEFRBs could just be such candidates.
If this hypothesis is true (and since the DMs of LEFRBs can exceed the prediction of the Galactic electron density model as shown in Figure \ref{FRB&RRATcomp}), we can make the following inferences: (i) the environment of the LEFRBs might be much denser than those of normal RRATs; (ii) the current Galactic electron density model might have missed some ionized gas; or (iii) LEFRBs locate in the Galactic halo and the halo might have a more significant DM contribution. 
In any case, Figure \ref{FRB&RRATcomp} also shows that the distribution of the DMs of RRATs could be naturally extended to the DM range of LEFRBs, although there seem to be a gap around $\rm DM\sim100\ pc~cm^{-3}$. 
In the future, a searching specially for sources around $\sim100\  \rm pc~cm^{-3}$ will be very helpful for clarifying the relationship between RRATs and LEFRBs.

\begin{figure}[htbp]
        \centering
        \includegraphics[width=0.45\textwidth]{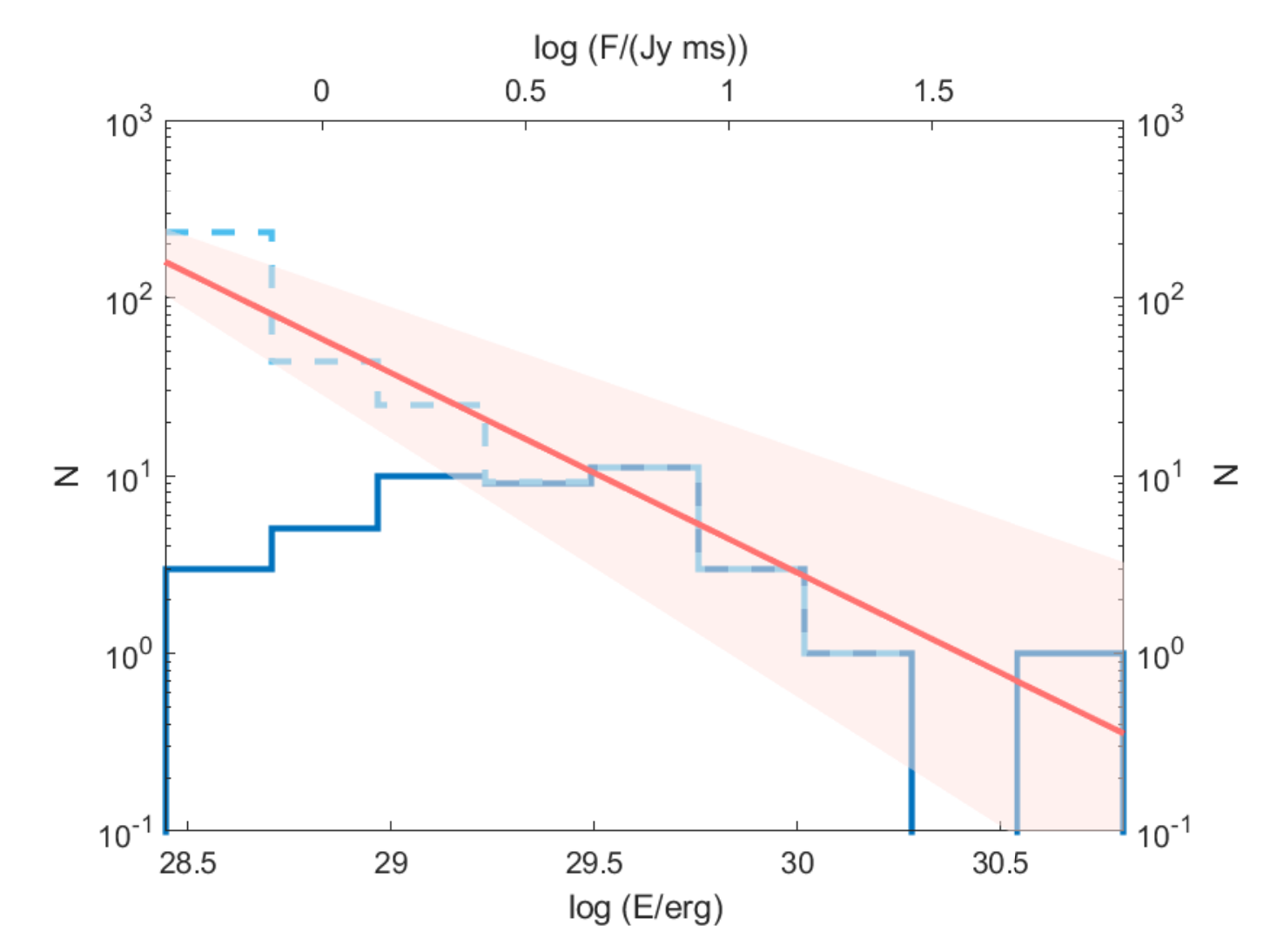}
        \caption{Energy distributions of LEFRBs without (blue solid) and with (blue dashed) a correction of the CHIME selection, where a reference distance of 10 kpc is used.
        The red solid line gives the best fit of the corrected distribution, which has a slope of -1.1. The shaded band represents the 95\% confidence range of the fit. 
}
        \label{LEFRB_Edis}
\end{figure}

In view of the potential Galactic origin of LEFRBs, we cannot estimate their distances by their DMs, since the DMs could be primarily contributed by the LEFRB sources themselves. 
Then, by taking a reference Galactic distance of $d=10$ kpc, we can recalculate the isotropically-equivalent energies of LEFRBs, which are in the range of $\sim 10^{28}-10^{31}$ erg as presented in Figure \ref{LEFRB_Edis}. 
These energies are much lower than that of the Galactic FRB 20200428 around $\sim10^{34}-10^{35}$ erg. 
Using Eq. (\ref{selection}), we correct the energy distribution of LEFRBs in Figure \ref{LEFRB_Edis} and find that the resultant distribution can be well fitted by a power law of an index $-1.1$, which is much flatter than that of HEFRBs. 
If we extend this power law to the energy range of FRB 20200428, then an extremely low event number would be obtained, which further indicates that LEFRBs are not likely to be the low-energy cousins of the magnetar-produced FRBs. 


\section{The event rate and origin of HEFRBs} \label{results}
A clue to the origin of HEFRBs can be inferred from the relationship of their redshift-dependent rates with the CSFRs, which has been investigated many times in previous works \citep[e.g.,][]{Cao2017a,Cao2018a,Zhang2021,James2022,Qiang2022,Zhang2022}. 
Here, we revisit this topic by taking into account the new factors discussed above, including the selection effects of CHIME, the independent determination of EF and the subtraction of the LEFRB samples. 
Specifically, we simulate the redshift and energy distributions of HEFRBs by using the Monte Carlo method as previously used \citep{Zhang2021,Zhang2022,Qiang2022}.
In short, we simulated mock FRBs with $(E,z)$ and use the selection effects Eq. \eqref{selection} to decide whether these mock FRBs can be ``observed" (or not).
The crucial inputs of this simulation are the energy $E$ and redshift $z$ probability densities of HEFRBs, which can be  described by Eq. (\ref{Enerdis}) and
\begin{eqnarray}
{p(z)}&\propto&\dot{R}_{\rm HEFRB}(z) \frac{1}{1+z} \frac{\mathrm{d} V}{\mathrm{d} z}\label{pz},
\end{eqnarray}
respectively, where $\dot{R}_{\rm HEFRB}(z)$ is the event rate of HEFRBs. 
Two representative types of event rates are considered as follows. 

\textbf{\textit{Case I:}} HEFRBs are produced by young NSs originating from the traditional channel of massive star core-collapses. 
Then, the redshift-dependence event rates of them are proportional to the CSFRs for the massive stars within the mass range of $(m_1,m_2)$. 
For normal NSs, we take $m_1=8\ M_{\odot}$ and $m_2=30 \ M_{\odot}$. 
Then, we have:
\begin{equation}
\dot{R}_{\rm HEFRB}(z) \propto f_{\rm m}\dot{\rho}_*(z),
\end{equation}
where the CSFRs are given by \citep{Yuksel2008},
\begin{equation}
\dot{\rho}_*(z)\propto\left[(1+z)^{a \eta}+\left(\frac{1+z}{B}\right)^{b \eta}+\left(\frac{1+z}{C}\right)^{c \eta}\right]^{1 / \eta},
\end{equation}
with $a=3.4, b=-0.3, c=-3.5, B \simeq 5000, C \simeq 9$, and $\eta=-10$. The fraction due to the mass requirement of the progenitors is expressed as:
\begin{equation}
f_{\rm m}={{\int_{m_1}^{m_2} \xi(m,z) \mathrm{d}m}\over {\int^{m_{\max}}_{m_{\min}} m\xi(m,z) \mathrm{d}m}},
\end{equation} 
%
where the initial mass function (IMF) of stars is adopted as \citep{Dave2008}:
\begin{equation}
\xi(m,z) \propto  
    \begin{cases}
        m^{-1.3}& \text{\emph{m}  <  $\hat{m}(z)$}\\
        m^{-2.3}& \text{\emph{m}  $\ge$  $\hat{m}(z)$}
    \end{cases}.
    \end{equation}

As a general consideration, we assume the break of the IMF could evolve with redshift as
$\hat{m}(z) = 0.5(1+z)^{\beta} M_{\odot}$, which leads to an extra evolution of the HEFRB rates. Such a consideration was previously adopted by \cite{Wang2011} to explain the redshift evolution of long gamma-ray bursts. 


\begin{figure}
    \centering
    \includegraphics[width=0.45\textwidth]{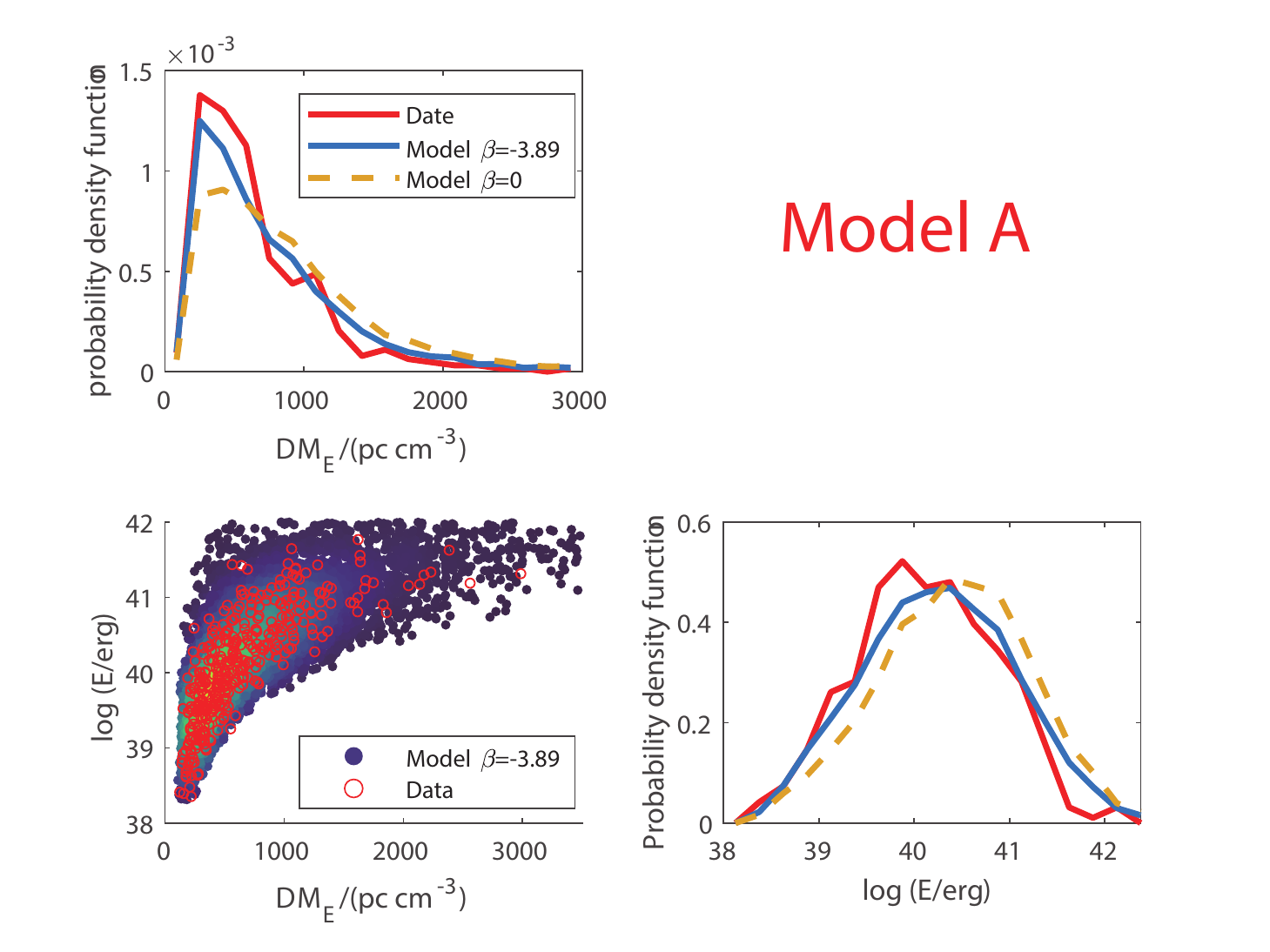}\\
    \includegraphics[width=0.45\textwidth]{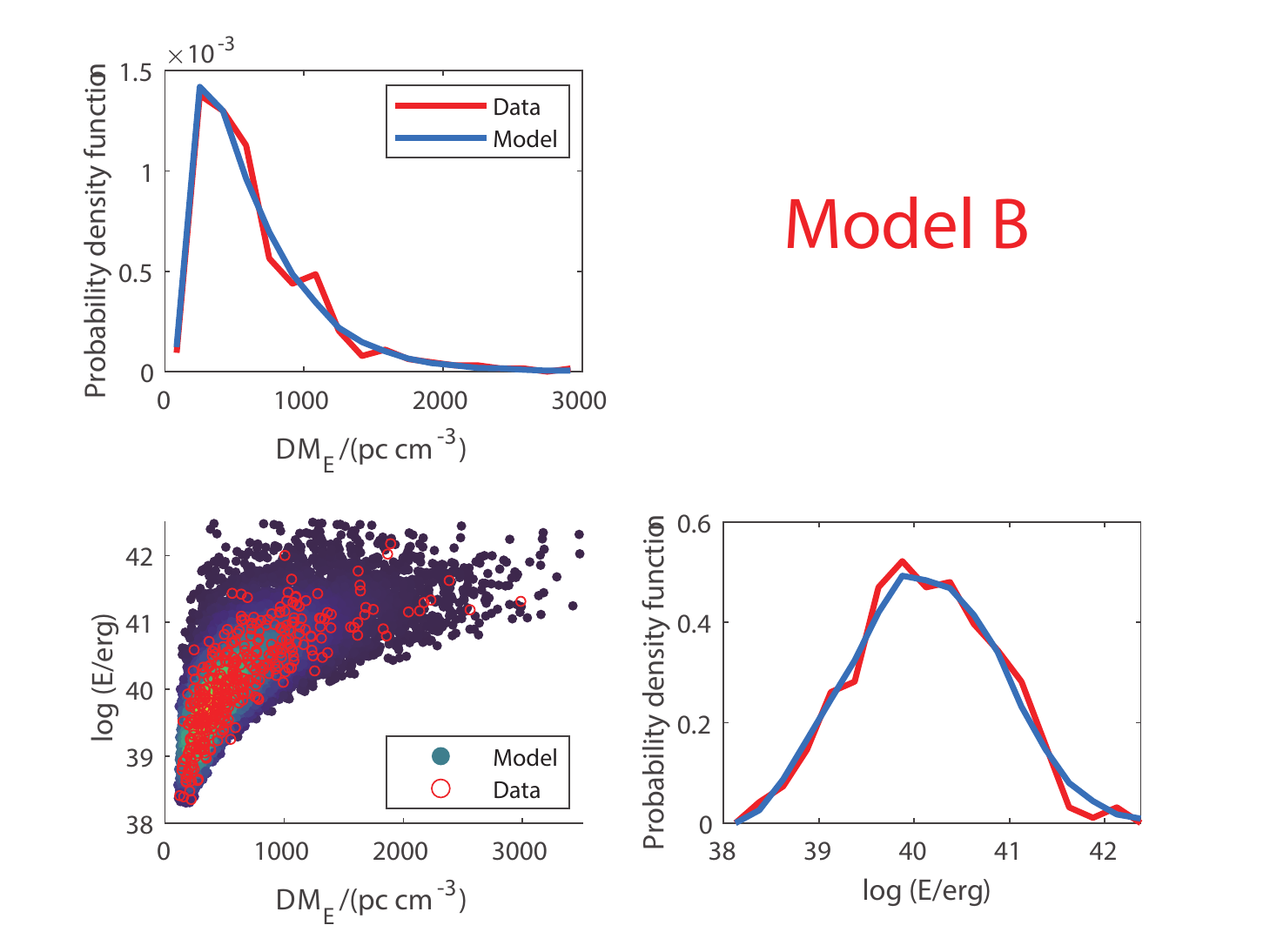}\\
    \includegraphics[width=0.45\textwidth]{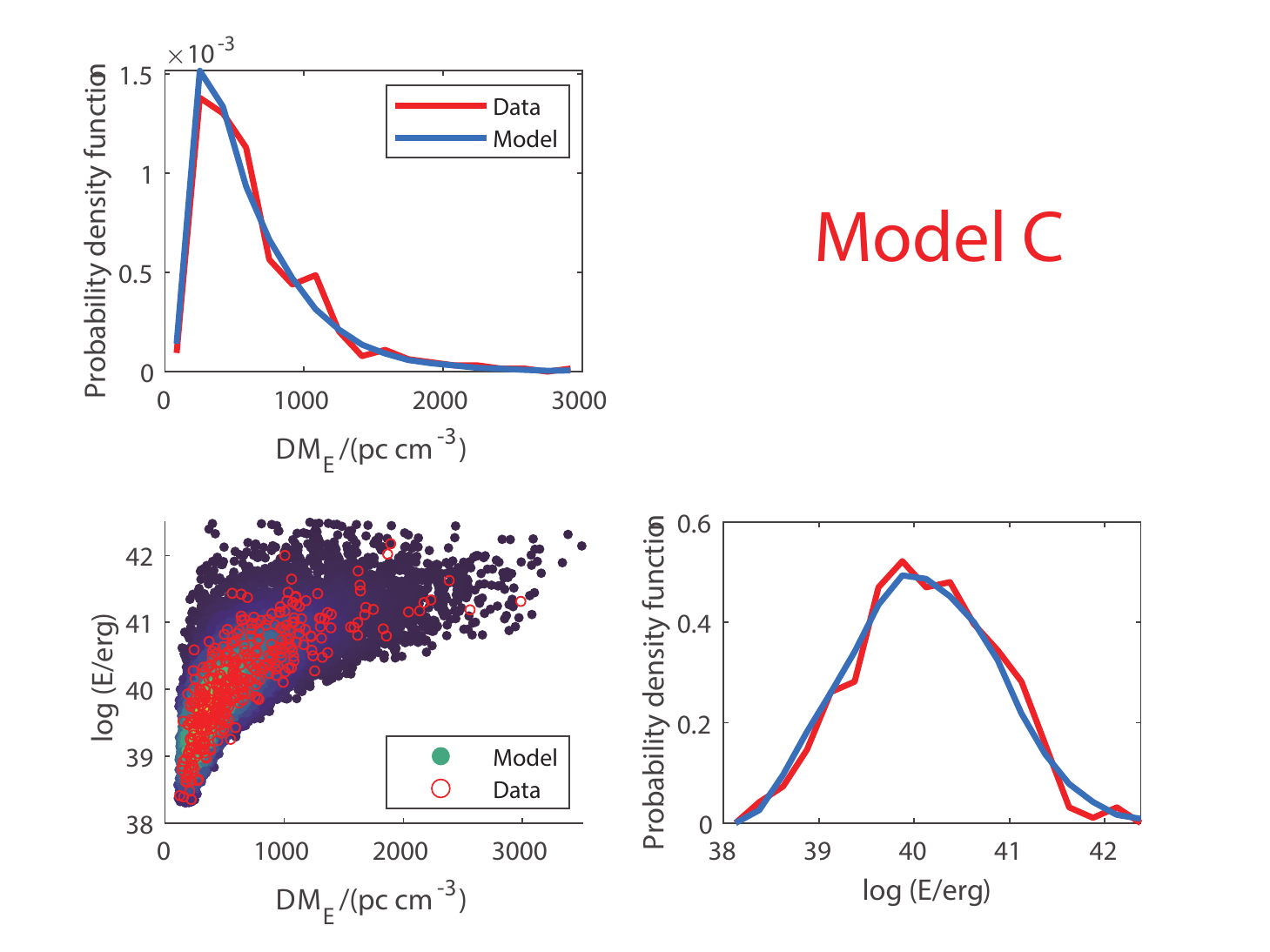}
    \caption{Comparison of the MC mock samples with the observational data of HEFRBs. For each model, the top left and bottom right panels show the $\rm DM_{E} = DM_{IGM} + DM_{host}/(1+\emph{z})$ and energy distributions, respectively.
    }
    \label{FitModel}
\end{figure}

\textbf{\textit{Case II:}} The young NSs producing HEFRBs are formed from the mergers of compact binaries (e.g., double NS, a NS plus a WD, and double WD, etc). 
In this case, the HEFRB rates are connected with the CSFRs by a delay time $\tau$ as \citep[e.g., ][]{Regimbau2009,Zhu2013,Regimbau2015,Tan2020}
\begin{eqnarray}
\dot{R}_{\rm HEFRB}(z)&\propto&(1+z)\int_{0}^{t-t_{\rm b}}{\dot{\rho}_{*}(t-\tau)\over 1+z(t-\tau)}P(\tau)\mathrm{d}\tau,\nonumber\\
&\propto&(1+z)\int^{z_{\rm b}}_{z(t)}{\dot{\rho}_{*}(z')\over 1+z'}P(t-t'){\mathrm{d} t\over \mathrm{d}z'}\mathrm{d}z'\label{mergerrate},
\end{eqnarray}
where the cosmic time $t$ is related to the redshift $z$ by 
$t(z)=\int_z^{\infty}[(1+z')H(z')]^{-1}\mathrm{d}z'$, and $z_{\rm b}$ and $t_{\rm b}$ represents the redshift and cosmic time at which the binaries started to be formed.
We set $z_{\rm b}=8$ in this work. Then,
$P(\tau)$ is the probability distribution of the delay time, which is usually assumed to take the form of \citep{Zevin2022arXiv220602814Z,Luo2022arXiv220607865L}:

Lognormal function,
    \begin{equation}
        P (\tau) \propto \exp\left[-\frac{( \textrm{ln}\ \tau- \textrm{ln}\ \tau_{\rm LN})^2}{2\sigma_{\rm LN}^2}\right].
    \end{equation}
    
    Power-law function,
    \begin{equation}
         P (\tau) \propto \left(\frac{\tau}{\tau_{\rm c}}\right)^{-1} e^{-\tau_{\rm c} / \tau} .
    \end{equation}   



\begin{table}[htbp]
\centering\caption{Parameters of HEFRBs}
\begin{tabular}{cccc}
\hline
\hline
Model & Parameter                           & $p_{{\rm KS},E}$ & $p_{\rm KS,DM_{E}}$ \\ \hline
A     & $\beta=-3.89$                       & 0.0072           & e-4                 \\ \hline
B     & $\tau_{\rm LN} = 0.23 \ \rm Gyr $   & 0.98             & 0.87                \\  
      & $\sigma_{\rm LN} =  1.52\ \rm Gyr$  &                  &                     \\ \hline
C     & $\tau_{\rm c} = 0.29 \ \rm Gyr$     & 0.82             & 0.25                \\ \hline
\end{tabular}
\label{table result}
\end{table}

The comparison of the mock samples with the observational HEFRBs are presented in Figures \ref{FitModel} for three different models, including the cases of core-collapse origin with an evolving IMF (Model A), merger origin of a lognormal distributed time delay (Model B), and merger origin of a power-law distributed time delay (Model C). 
The parameters of the best fits and the corresponding $p_{\rm KS}$ values of these models are listed in Table \ref{table result}. 
First of all, as previously found by \cite{Cao2017a} and \citet{Zhang2022}, our Figure \ref{FitModel}(a) shows that the core-collapse origin model  obviously deviates from the observational distributions, no matter whether an evolving IMF is invoked or not. 
Instead, as we show in Figures \ref{FitModel}(b) and \ref{FitModel}(c), the fitting of the distributions can be improved significantly by invoking a time decay, for both the lognormal case and power-law case.
The constrained values of the characteristic delay time are basically consistent with the results of \citet{Cao2018a} and \citet{Zhang2021}, but smaller than that in \citet{Zhang2022}. 
We can only roughly investigate which compact star mergers correspond to this characteristic delay time by binary population synthesis, because there are uncertainties in population synthesis.
Considering Roche-lobe overflow and common-envelope evolution in both single-degenerate scenario and double-degenerate scenario of type Ia supernova, \citet{Mennekens2010} gave a delay time distribution that peaks in around 0.1 Gyr for double WD mergers.
Taking into account wide range initial conditions and different NS natal kick distributions, \citet{Toonen2018} found that the mergers of NS and WD typically explode in $0.01-1$ Gyr.
cRecently, \citet{Kobayashi2022} showed that previous binary population synthesis models are hard to explain the observed elemental abundances in the Milky Way and a shorter delay time distribution around 0.1 Gyr is needed for double NS mergers.
Roughly speaking, our characteristic delay time around $0.2-0.3$ Gyr is close to the delay times of NS-WD, double NS, and double WD mergers.

Finally, We can estimate the local event rate of HEFRBs as follows. 
In order to get 383 detectable HEFRBs of distributions same to the observations, we need to generate a total of about 61,000 mock samples, among which about 181 samples locate within the distance of 1 Gpc. 
Therefore, the local event rate of HEFRB can be estimated to be:
\begin{eqnarray}
\dot{R}_{\rm HFRB}(0)&\sim&\dot{R}_{\rm sky}\rm \cdot{380HEFRB\over 600FRB}  \cdot{61000\over 380}\nonumber\\
&&\rm \times{181~Gpc^{-3}\over 61000~sky^{-1}}\cdot{ 365day\over yr}\nonumber\\
&=&9\times10^{4} \rm \ Gpc^{-3} yr^{-1},
\end{eqnarray}
where $\dot{R}_{\rm sky}\sim 820 \ \rm sky^{-1} day^{-1}$ is the full sky rate of FRBs given by \citet{CHIME2021}. This result is somewhat higher than the rate inferred from the Parkes observations \citep{Cao2018a} because the EF here can be extended to lower energy represented by $E_{\rm c}$.


\section{Summary and conclusions} \label{conclusions}
The origin of FRBs is one of the biggest mysteries in current astronomy, which might have various different answers. 
For example, different origins are very likely to be owned by the repeating and non-repeating FRBs, even though the latter ones could also be intrinsically repeating on sufficiently long timescales. 
In this paper, we investigate the statistical distributions of the non-repeating FRBs detected by CHIME and we find that, even for these similar bursts, there could be different generation channels, as implied by the apparent low-energy excess in the energy distribution. 
In particular, we find that LEFRBs are concentrated towards the Galactic plane and their latitude distributions are similar to that of RRATs. 
These indications hint that the LEFRBs might compose a special type of RRATs characterized by only one detected pulse, unusual high DMs, and relatively high energies $\sim10^{28-31}$ erg.
One possibility is that the environments of these special RRATs are much denser than those of normal RRATs. 
For the remaining non-repeating cosmological FRBs, our statistical study demonstrates again that
they can be produced by remnant NSs of compact binary mergers, including mergers of double NS, a NS plus a WD, or double WD.

\begin{acknowledgements}
We thank Liang-Duan Liu, Chen-Hui Niu, Xia Zhou, and Yuan-Chuan Zou for helpful discussion and comments.
We thank the anonymous referee for improving the paper.
This work is supported by the National Key R\&D Program of China (2021YFA0718500), the National SKA program of China (2020SKA0120300), 
and the National Natural Science Foundation of China (grant Nos. 11833003 and U1838203).
MATLAB \citep{MATLAB2021} and MCMC tool box for Matlab \url{https://mjlaine.github.io/mcmcstat} are used in this work.
\end{acknowledgements}

\bibliographystyle{aa}
\bibliography{reference.bib}

\end{document}